\documentclass{article}

\usepackage{arxiv}

\usepackage[utf8]{inputenc} 
\usepackage[T1]{fontenc}    
\usepackage{hyperref}       
\usepackage{url}            
\usepackage{booktabs}       
\usepackage{amsfonts}       
\usepackage{nicefrac}       
\usepackage{microtype}      
\usepackage{lipsum}
\usepackage{graphicx}
\graphicspath{ {./images/} }

\title{Dispersion Study of a Broadband Terahertz Focusing Reflecting Metasurface for 6G Wireless Communication}

\author{
Fahim Ferdous Hossain\\
  School of Electrical and Computer Engineering\\
  Oklahoma State University\\
 Stillwater, OK 74078, USA\\
  \texttt{fferdou@okstate.edu} \\
   \And
John F. O'Hara \\
  School of Electrical and Computer Engineering\\
  Oklahoma State University\\
 Stillwater, OK 74078, USA \\
  \texttt{oharaj@okstate.edu}
}

\begin{document}
\maketitle
\begin{abstract}
In 6G wireless communications, functional terahertz reflecting metasurfaces are expected to play increasingly important roles such as beamforming and beamsteering.  This paper demonstrates the design of a functional and efficient beamforming metasurface in the burgeoning D-band (0.11-0.17~THz). In addition to achieving broadband operation (0.135-0.165~THz), this design is polarization-maintaining, diffraction limited, simple in design, exhibits 64.1\% broadband efficiency (1.9 dB insertion loss) and 20\% fractional bandwidth. Despite being formed by an array of highly dispersive resonators, the metasurface exhibits very low temporal dispersion, which avoids pulse reshaping and its consequent limitations on achievable data rate. The design and performance of the focusing reflector are presented followed by a group delay and group delay dispersion analysis revealing that a 2.83\% temporal broadening of the pulse is observed at the focus.  
\end{abstract}


\section{Introduction}
Next-generation (6G) wireless communication technology will require significantly higher data rates and more tailored waveform design than 5G technology \cite{Akyl_6G_Beyond(1)}. The higher data rate requirement demands the utilization of the terahertz frequency band \cite{Elme_6G_Wire(2)}, pushing carrier frequencies above 0.1~THz \cite{Kokko_LOS(3)} where large contiguous blocks of bandwidth are readily available.  Atmospheric attenuation is a key challenge in terahertz wireless communication \cite{Naga_Adv(4)}. However, at a given water vapor density, D-band frequencies suffer relatively manageable atmospheric attenuation, particularly compared to higher frequencies in the terahertz regime \cite{Kokko_LOS(3), Naga_Adv(4), Xing_Propa(5)}, and communication is now becoming possible into the multi-kilometer ranges \cite{Naga_Adv(4)}. Currently, if the the D-band is allocated for wireless communications, it can provide contiguous bands up to 32.5~GHz, bounded on either side by RR5.340 forbidden bands \cite{Coshare_2021}. As such, D-band links could feasibly be exploited to implement high data rate wireless backhaul\cite{Xing_Propa(5)}. 

With narrow and more easily scattered terahertz beams utilized in 6G communications, future wireless links will require improved control over the beam (e.g. steering, focusing, wavefront correction). Metasurfaces have the potential to become a major player, finding applications in smart radio environments, massive multiple input multiple output (MIMO) technology \cite{Ai_Multi(6),Ultra_MIMO_Jornet}, and reconfigurable intelligent surfaces (RIS) \cite{P_Marios}.  Metamaterials have marshaled a lot of interest over the past two decades by their ability to manipulate electromagnetic waves in an unprecedented manner \cite{Engheta_Meta_Phy(6)}. Metamaterial demonstrations include stealth \cite{Schurig_Meta_Cloak(7), Zhong_Radar(8)}, negative refraction\cite{Ling_Negat(9), Shelby_Experi(10)} and super lensing \cite{Wong_Optical(11)}. Planar metamaterials with subwavelength thickness are called metasurfaces and are regarded as the 2D counterpart of 3D metamaterials\cite{Chen_Review(11),Spatio_Temp_Shaltout, Holloway}. Metasurfaces are typically easier to fabricate, lighter, less bulky, achieve highly practical functionality \cite{Chen_Review(11)}, and can exhibit less loss than bulk metamaterials \cite{Holloway}. This would be particularly important in space-based applications.  Metasurfaces have been demonstrated from the microwave to visible regimes including terahertz frequencies \cite{Chen_Review(11)}. Demonstrated terahertz metasurface devices already include: planar convex/concave metalenses, holograms, wave plates,  beam splitters, special beam generators, arbitrary polarization controllers, and various active devices \cite{Zang_Manip(12)}.  

Several specific examples illustrate the state of the art in terahertz metasurfaces. The design of a metasurface mirror with adaptive focusing operating in the vicinity of 2~THz was presented by Hosseininejad et~al. \cite{Hossein_Repro(14)}. In \cite{Gu_Effi(15)}, a small, efficient terahertz focusing lens with 4.7~mm focal length was presented. Lee et~al. experimentally demonstrated a quarter wave mirror based on dielectric resonators operating between 0.97\textasciitilde1.6 THz having a fractional bandwidth of 49\% and reflection amplitudes exceeding 92\% for both orthogonal polarizations\cite{Lee_Dielres(16)}. In reference \cite{Lee_Dielres(16)}, a half wave mirror also based on dielectric resonators operating in the 0.89\textasciitilde1.54 THz frequency range has been experimentally demonstrated, having a peak/average cross-polarization reflection amplitude greater than 72\%/ 79\%, respectively, and a peak/average co-polarization reflection amplitude smaller than 31\%/15\%.

State-of-the-art focusing metasurfaces can be better illustrated by considering wavelengths in addition to terahertz. A metasurface based lens with focal distances of 3~cm and 6~cm was demonstrated using a flat axicon and operated at the telecom wavelength (1.55~$\mu$m) \cite{Aieta_Aberfree(17)}. General strategies to design meta-mirrors employing a Lorentzian multi-resonance model operating in microwave (8-12~GHz), terahertz (0.5-0.8~THz), and optical (1,150-1,875~nm wavelength) ranges have been proposed in \cite{Tang_Ultrawide(18)} with experimental demonstration of an achromatic and abnormal chromatic meta-mirror in the 8-12~GHz range. A terahertz focusing mirror with super-resolution capability consisting of unit cells having reflection magnitude more than 88\% operating in the 200-300~GHz frequency range has been proposed in \cite{Wang_Broad}. One optical metalens operating between 470-670~nm was designed by utilizing nanofins on a surface \cite{Capsso_Achrom_Visi(20)}. 

Because of their usual resonant and frequency-dispersive nature, metasurfaces commonly present challenges to achieving broad bandwidth, particularly in the terahertz regime, where fewer constituent materials are available for complex structural design, compelling additional research. In \cite{Witha_Enhanced_(21)}, a terahertz reflectarray operating with a center frequency of 1~THz was shown to have a  fractional bandwidth of about 24\%. An impressive experimental demonstration of a broadband achromatic terahertz metalens having 91\% fractional bandwidth in the 0.3-0.8~THz regime was also presented \cite{Qingqing_achro(22)}, though this approach relied on a complex all-dielectric deep-etching fabrication. It is finally noted that some communication and sensing applications may benefit from longer focal length designs, particularly those that do not alter or rely on wave polarization effects. The design of a polarization-preserving, long focal length, focusing metasurface reflector, having simple design, high broadband efficiency, and wide bandwidth has not been yet demonstrated in the D-band frequency range. 

Because of the ubiquitous need to utilize wireless spectrum efficiently, the dispersive behavior for meta-reflectors warrants particular attention, especially as operating bandwidths become large, as in 6G. For example, terahertz waves undergo group delay dispersion (GDD) while propagating through the atmosphere due to the frequency-dependent refractivity of air\cite{Karl_Comp(7),Hill_Disp(8),Mandeh_Exp(9)}.  This can occur to the extent that it causes intersymbol interference (ISI), which leads to reduced available data rates \cite{Karl_Comp(7),Karl_Funda}. While GDD has traditionally been addressed in wireless systems by digital equalizers, their practical implementation may not be optimal in light of the more challenging requirements of 6G waveforming, synchronization, noise, and baseband signal processing. Moreover, the effectiveness of digital equalization to mitigate the effects of GDD in 6G systems remains unvalidated \cite{Heath_over(10),Sari_over(11),Farrell_Perf(12)}, particularly as 6G waveforms continue to be developed. Metasurfaces can similarly produce GDD and it is therefore important to quantify the dispersion introduced to broadband communication waveforms by metasurface-based reflectors. Such dispersion would add to any existing dispersive elements of the channel (e.g. atmosphere) and therefore must be evaluated for its impact on 6G wireless systems and any possible imposed design constraints. 

In this paper, we propose a design of a flat, metasurface focusing reflector based on metal-insulator-metal (MIM) structured unit cells, having a relatively long design focal length of 500~mm, and operating in the 135-165~GHz range. The behavior of the reflector has been quantified in terms of GDD, pulse broadening, broadband focal length, optical focusing performance, and power efficiency. The reflector achieves nearly diffraction-limited focusing with a broadband focal length of 500~mm, a broadband power efficiency of 64.1\% and temporal pulse broadening of 2.83\% at the focus.  This particular case-study was designed as a stepping stone toward improving the performance of long-path terahertz spectroscopy cells \cite{Tae_N20, Tae_water}, however the study of its dispersive behavior is equally informative for applications like 6G.

\section{Design of focusing reflector}
To minimize losses, the metasurface reflector design should aim for a unity reflection coefficient magnitude over its entire surface area.  Assuming an incident plane wave, the phase distribution of the wave reflected from the flat 2D metasurface should satisfy the following equation if the reflector is assumed to be in the $xy$-plane and its center is coincident with the origin of the three-dimensional Cartesian coordinate system:
\begin{equation}
\Delta\phi=\left(\frac{2\pi}{\lambda}\sqrt{(x^2+y^2+f^2)}-f\right)mod 
  \ 2\pi,
\label{eqn:1}
\end{equation}

  \noindent Here, $\Delta\phi$ is the phase difference required at the transverse point $(x,y)$ on the metasurface relative to the phase at the origin, $\lambda$ is the wavelength and $f$ is the focal length of the reflector. The optical axis is assumed to be coincident with the $z$-axis. The modulo $2\pi$ in Eq.~\ref{eqn:1} ensures that phase differences greater than $2\pi$ are adjusted to remain between $0$ and $2\pi$. The task of the metasurface is to produce this phase profile on the wave, along with nearly complete reflectivity, upon reflection. The frequency and spatial dependency of this phase specification reveal that multiple parameters must be simultaneously tuned in the final metasurface design. And since this profile is continuous in space, the discretized nature of metasurface elements will necessarily limit the achievable fidelity to a step-wise approximation.
\begin{figure}[h!]
    \centering
    \includegraphics[width= 13 cm]{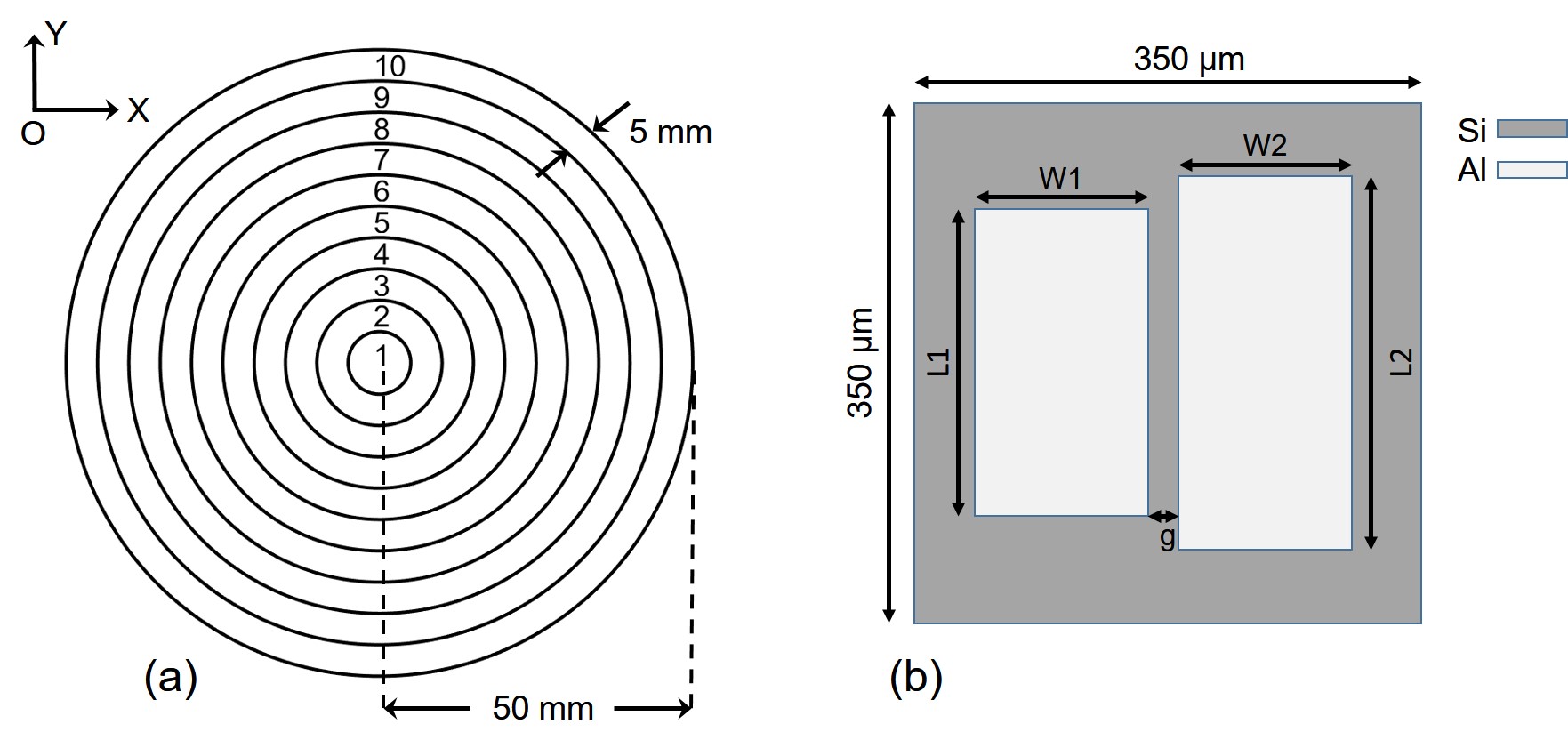}
    \caption{a) A schematic of the focusing reflector where the digits 1 to 10 correspond to individual annular regions. The radius (50~mm) of the reflector and width of a region (5~mm) is indicated. The XY arrowset  denotes that the reflector surface is situated in the XY plane. b)A sample unit cell of MIM structure. $L1$, $W1$, $L2$, $W2$, and $g$ were varied to tune the reflection response. The metal backplane cannot be seen in figure.}
    \label{fig:unit_cell}
\end{figure}

\noindent To approximate the phase relationship in Eq.~\ref{eqn:1}, the metasurface reflector proposed in this work consists of nine annular regions, each having a width of 5~mm, plus a single 10~mm diameter circular region in the middle, corresponding to a total lens diameter of 100~mm. A schematic of the focusing reflector is depicted in Fig~\ref{fig:unit_cell} (a). Each of these regions has been populated by many instances of a corresponding unit cell. The unit cells are MIM structures designed by inspiration from \cite{Azad_Meta_MW}, and the sample structure can be viewed in Fig~\ref{fig:unit_cell} (b): two rectangular aluminum patches on a 50~$\mu$m thick silicon substrate with an aluminum backplane. The side length of the square unit cell was 350~$\mu$m. The geometric parameters of the rectangular aluminum resonators such as length, width, and gap between resonators were varied to achieve a 0 to 2$\pi$ phase range required in the reflected wave. COMSOL Multiphysics simulation software was used to obtain the complex reflection coefficients of the unit cells utilizing the finite element method (FEM). Numerous unique unit cells were simulated and ten were judiciously chosen based on the phase, amplitude, and frequency-dependence of their reflection coefficients. 

\begin{figure}[ht!]
    \centering
    \includegraphics[width= 13 cm]{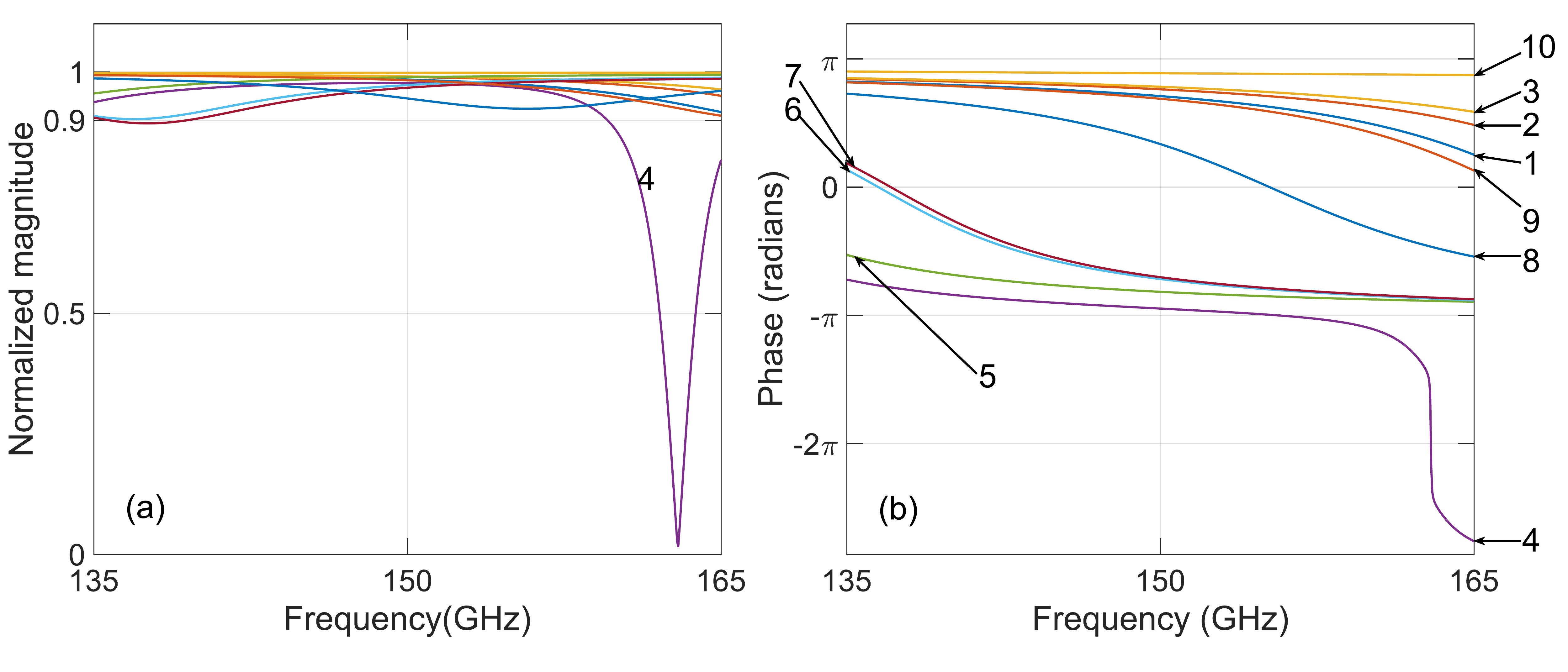}
    \caption{a) Magnitude of the reflection coefficient for the selected unit cells plotted with respect to frequency. The digit 4 denotes the curve corresponding to unit cell 4. b) Reflected wave phase for the selected unit cells plotted with respect to frequency. The digits at the ends of the arrowheads denote indices of the unit cells.}
    \label{fig:mag}
\end{figure}

The reflection coefficient magnitudes of the unit cells are presented in Fig.\ref{fig:mag} (a). All the unit cells except those in region 4 show reflection coefficient magnitudes exceeding 0.9 over the entire design bandwidth. The reflection coefficient magnitude curve of unit cell 4 shows a strong resonance between 161-165~GHz, dramatically reducing its efficiency in this range. It may seem that the majority high values of reflection coefficient magnitudes will translate into high broadband device efficiency too. However, in practice imperfect interference at the focus also plays a role, so that a holistic investigation of the reflector response is required. This is best investigated after the metasurface is known to function as an effective focusing reflector. 

\begin{figure}[h!]
    \centering
    \includegraphics[width= 12 cm]{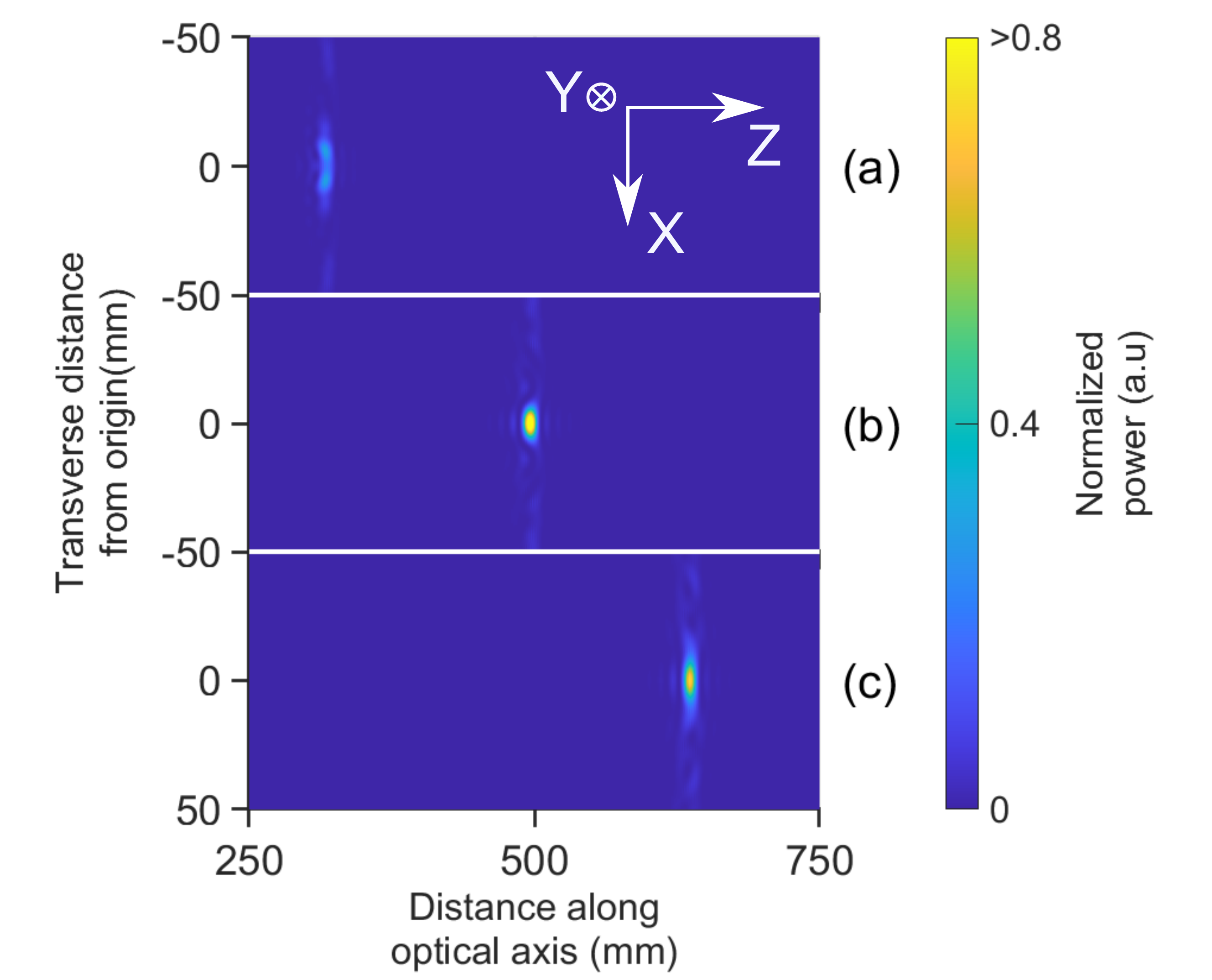}
    \caption{Broadband pulse reflected from the metasurface reflector at three different time instances. The instant when the pulse was incident on the reflector was considered as time zero. a) Broadband pulse before focal plane (1.07~ns), b) pulse at the focal plane (1.67~ns) and c) pulse after the focal plane (2.13~ns). According to the coordinate system given in a), the reflector is situated in the $xy$-plane centered at the origin and the pulse is propagating along the optical ($+z$) axis. Colorbar labels are normalized to peak intensity at focus.}
    \label{fig:pulse_focus}
\end{figure}

\noindent The focusing behavior of the designed metasurface can be approximated with a Huygens-Fresnel treatment. Each metasurface unit cell is considered a point radiator that is initially stimulated by illumination from the incident broadband wave whose phase fronts are parallel to the $xy$-plane. The waves are modified by the complex reflection response of the various unit cells as a function of position and thereby produce an approximation to the phase profile given in Eq.~\ref{eqn:1} and the requisite constructive interference near the focal region of the reflector. Since the incident wave is a transform-limited broadband pulse, the reflected wave is also localized in space and time and can be visualized at different time instances, as shown in Fig.~\ref{fig:pulse_focus}. Here, three different time instances of a 30~GHz broadband reflected pulse are depicted.
\noindent The plots reveal several important attributes about the performance of this metasurface reflector. First, it achieves nearly diffraction-limited focusing, as seen in Fig.~\ref{fig:pulse_focus}(b), where the transverse, ($1/e^2$) beamwidth of the focus is approximately 14.6~mm. This matches well with the Rayleigh-Sommerfeld diffraction theory \cite{Goodman} estimate of 13.8~mm from an ideal reflector. Second, the integrated power contained in the focused pulse reveals that the overall efficiency of the metasurface reflector is 64.1\%, when normalized to an ideal, concave, focusing mirror. Third, the narrow and single-peaked shape of the pulse in the $z$-direction reveals that very little overall temporal broadening occurred, despite the dispersive nature of the metasurface unit cells. This was by design, of course, but raises the issue of exactly how well, quantitatively, the design actually worked. To answer that, the concept of dispersion must be elaborated.


\begin{figure}[h!]
    \centering
    \includegraphics[width= 13 cm]{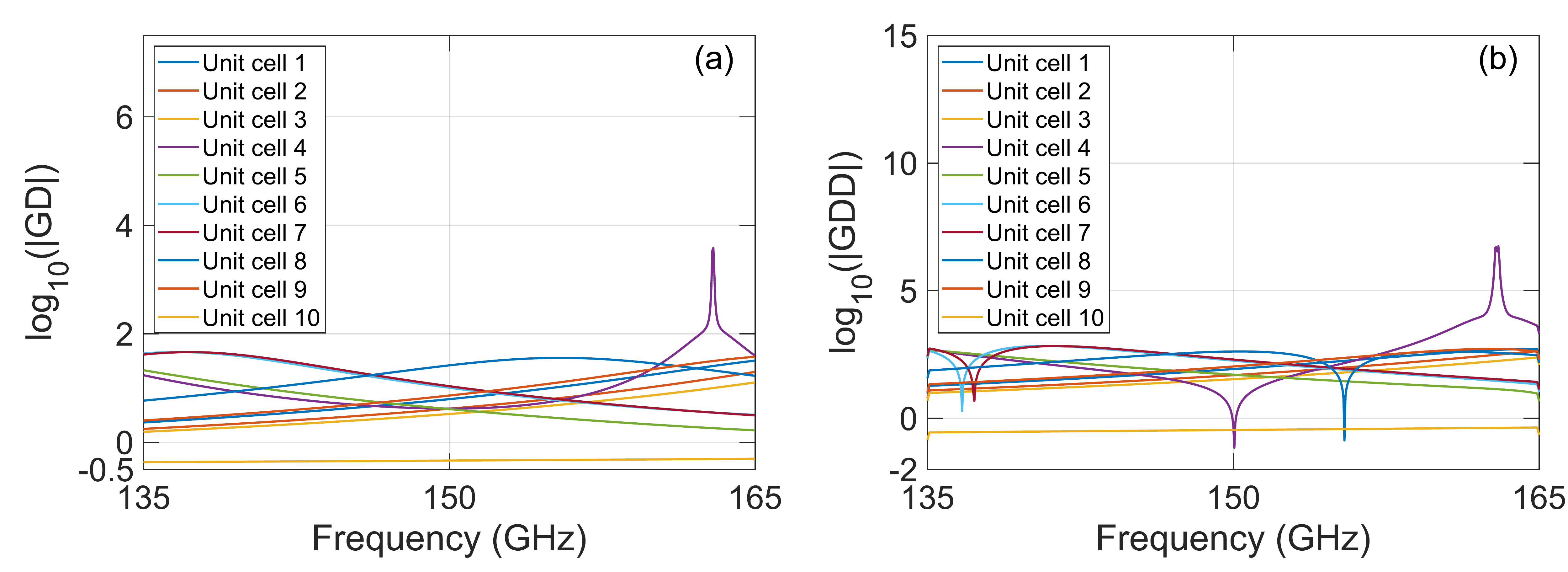}
    \caption{a) Group delay variation with respect to frequency for selected unit cells plotted in log scale. The unit of GD is picoseconds. b) Group delay dispersion variation with respect to frequency  for selected unit cells plotted in log scale. The unit of GDD is ps$^2$. The magnitude log-scale is used to clarify the GD and GDD behavior.}
    \label{fig:GDD}
\end{figure}
\section{Results and Discussion}
Dispersion effects may be quantified by starting with group delay, which is defined as\cite{Oppenheim}:

\begin{equation}
 GD=-\frac{d\phi}{d\omega}
 \label{eqn:2}
\end{equation}

\noindent where $\omega$ is the angular frequency and $\phi$ is the phase of the wave reflected by the metasurface.  Extending this principle, group delay dispersion (GDD) is defined with another derivative \cite{Russ} or 

\begin{equation}
GDD=-\frac{d^2\phi}{d\omega^2 }.
\end{equation}

\noindent According to Eq.~\ref{eqn:2}, a linear phase-frequency relationship entails a constant GD and zero GDD.  This represents the ideal case when designing a focusing broadband reflector, thus two main design parameters in the phase must be considered. First, to ensure focusing at the correct focal length, each unit cell or annular region of the reflector must satisfy Eq.~\ref{eqn:1} for all the involved frequencies. Second, to ensure zero GDD, the phase response at any spatial location on the reflector must be linear as a function of frequency. The phase-frequency relationship of the chosen unit cells for our design is shown in Fig.~\ref{fig:mag}(b) over a frequency range of 135-165 GHz. From our available pool of simulated unit cells, the chosen unit cell set complied most favorably with both aforementioned phase requirements and with the requirement to maximize reflection magnitude. However, from Fig.~\ref{fig:mag}(b), it is evident that the phase–frequency relationships of the unit cells are not linear over the given frequency range, especially for the unit cells in region 4. Any such nonlinearity causes the reflector response to have a frequency dependent group delay, which is shown for each unit cell design in Fig.~\ref{fig:GDD}(a). Each of the reflector regions shows variable group delay over the frequency range of interest. The unit cell chosen for annular region 4 shows a large peak in group delay in the 162.5-163.5 GHz range. This can be attributed to the unit cell's resonance in this frequency range, which is accompanied by a sharp change in phase. With few exceptions, group delays for all other unit cells and for unit cell 4 in most of its frequency range are less than 46~ps.
\begin{figure}[h!]
    \centering
    \includegraphics[width=\columnwidth]{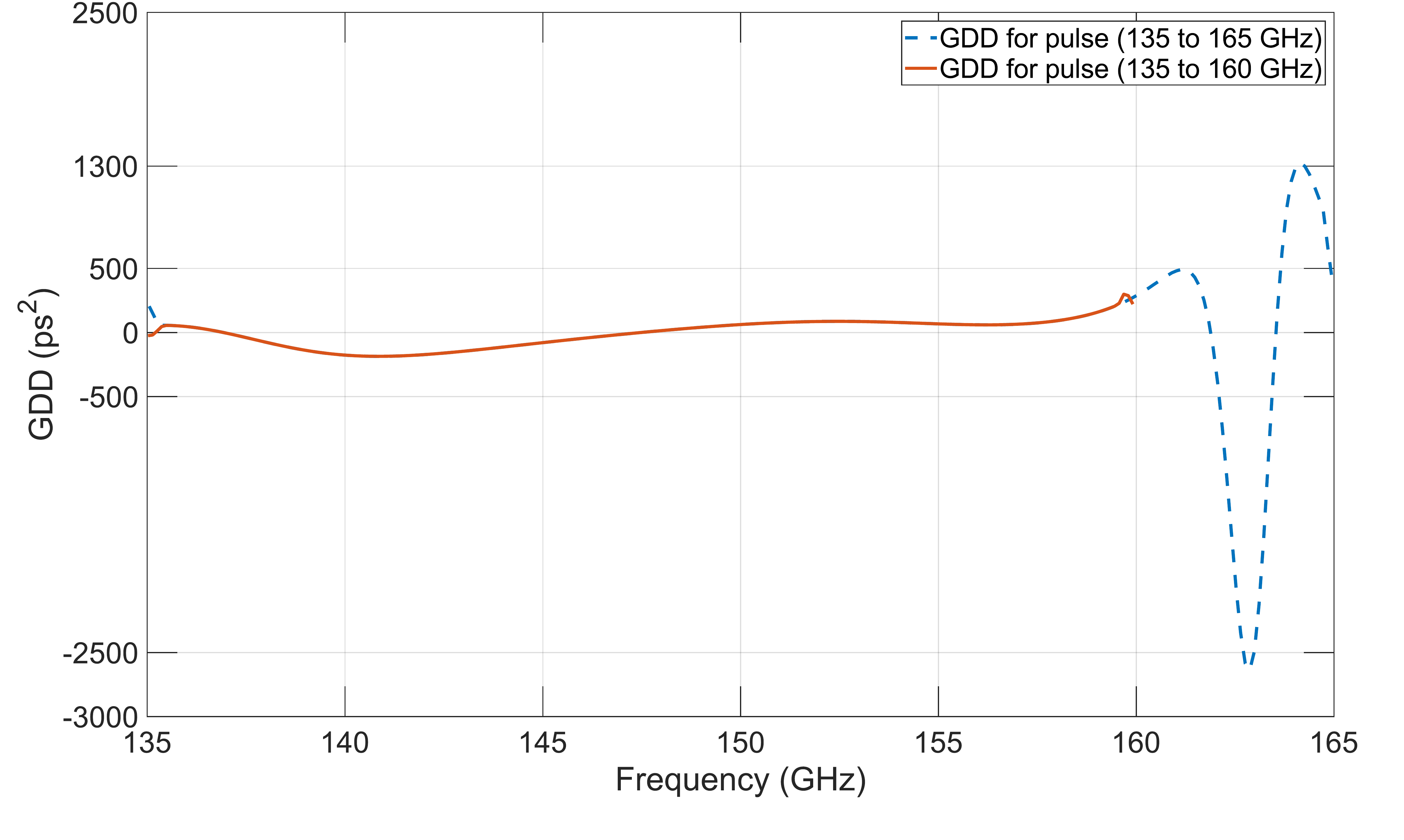}
    \caption{ Group delay dispersion computed for the focused pulse of the metasurface focusing reflector with bandwidth of 135-165~GHz (blue dashed) and 135-160~GHz (red solid).}
    \label{fig:GDD_Focus}
\end{figure}

GDD indirectly quantifies how much temporal dispersion a broadband pulse experiences. To maintain the temporal duration of a broadband pulse, as would typically be desired in wireless communications, low values of GDD are desired. Anywhere group delays of unit cells are frequency dependent, non-zero GDD is expected. As such, the GDD imposed by reflections from region 4 should be most impactful.  In Fig.~\ref{fig:GDD}(b), the GDD behavior with respect to frequency can be observed. Mostly, the values of GDD are within or close to $-680$-545~ps$^2$. However, in region 4, GDD spikes to values exceeding $\pm 5\times10^6$~ps$^2$, over a narrow bandwidth. Such large values hint at the possibility of observing pulse broadening in the time domain. However, the fields at the focus integrate the effects of all the regions of the reflector, meaning the GDD effects from region 4 will be dampened in the overall performance. 

\begin{figure}[h!]
    \centering
    \includegraphics[width=\columnwidth]{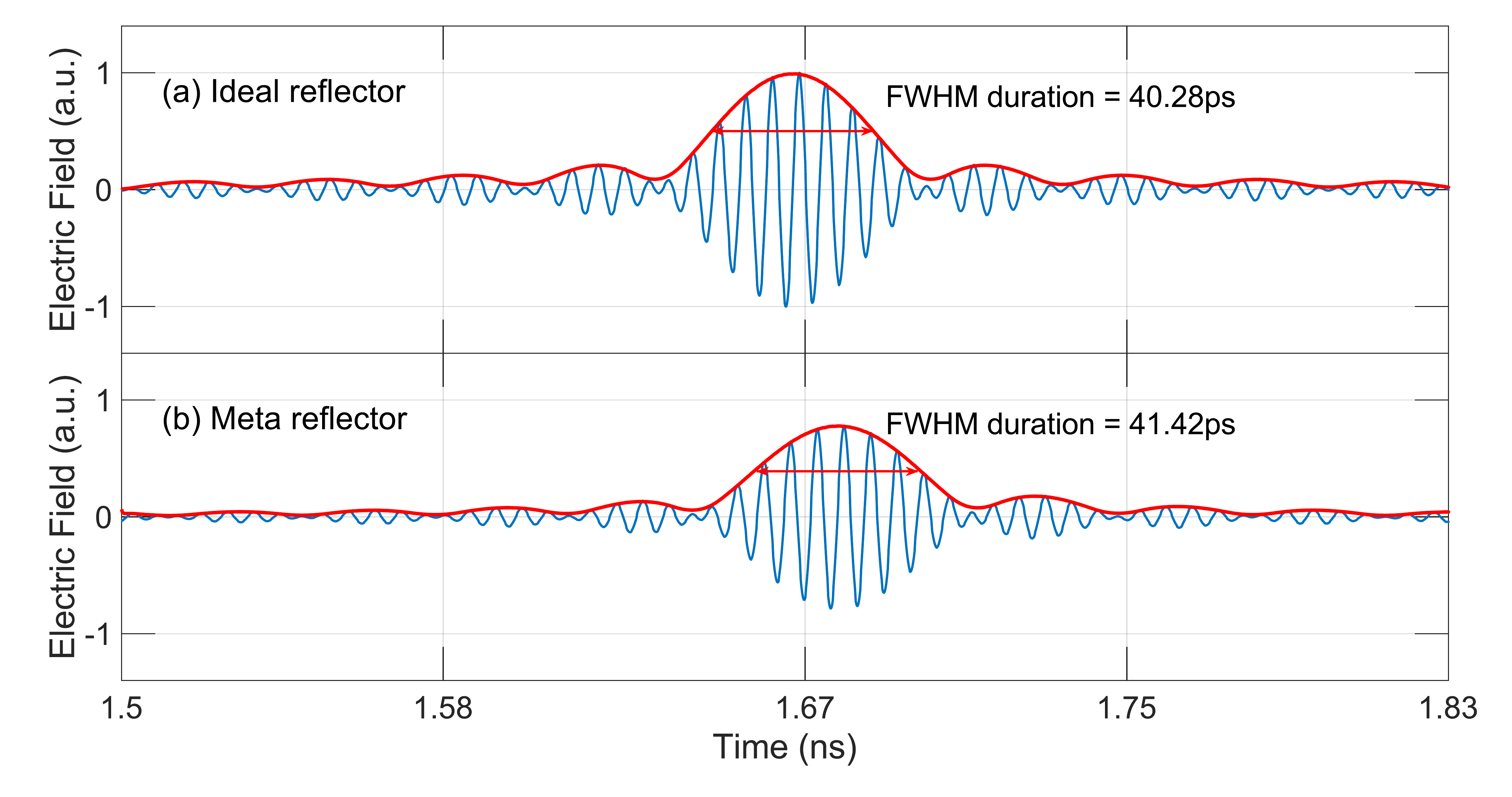}
    \caption{a) Broadband pulse with 30~GHz bandwidth (135-165~GHz) focused by an ideal reflector normalized to its maximum field value b) broadband pulse focused by the metasurface reflector normalized to the maximum field value of the pulse focused by the ideal reflector. The red curves are envelopes of the pulses and the horizontal line segments indicate FWHM pulse widths in space. The starting value of the time axis is arbitrary. }
    \label{fig:Pulse_at_Focus}
\end{figure}


The GDD of the pulse at the focus is presented in Fig.~\ref{fig:GDD_Focus}. Here, GDD values stay mostly between $-190$ to $190$~ps$^2$ with expected peaks at the band edges and one large feature around 162-164~GHz. This GDD behavior is a combined effect of imperfect focusing of individual regions of the reflector (due to imperfect GD values) and their non-zero GDD values, particularly those in region 4. In the time domain, this GDD behavior also corresponds to a broadening of the pulse at the focus. The input pulse incident on the reflector had a FWHM duration of 40.28~ps, and this was unchanged at the focus when an ideal mirror was substituted for the metasurface reflector. However, the pulse at the focus of the metasurface reflector had a FWHM duration of 41.42~ps, a 2.83\% increase in the pulse width, as shown in Fig.~\ref{fig:Pulse_at_Focus}. In a communication system, this increased pulse width proportionally translates to a lower spectral efficiency and consequently a lower achievable bit rate, if left uncorrected. An incident pulse having reduced bandwidth of 135-160~GHz was also studied to observe the effect of removing the GDD spike of region 4. The corresponding GDD curve is shown in Fig.~\ref{fig:GDD_Focus} and that depicts an almost complete match with the original GDD curve. In this reduced bandwidth case, the incident pulse's FWHM duration was 48.26~ps and the FWHM duration at the focus was 48.83~ps which results in a pulse width broadening of only 1.18\%. The reduced impact of pulse broadening in this case illustrates the significant impact of the GDD spike of region 4, which resulted in a 140\% pulse broadening penalty.

The power efficiency of the metasurface reflector compared to an ideal mirror is 64.1\%, despite all but one of the metasurface unit cell designs exhibiting reflection coefficient magnitudes $>0.9$ for the entire operating bandwidth. The reason for can be attributed to the imperfect group delays. The power efficiency was computed as 
 $\Sigma I_{\mathrm{foc}}/\Sigma I_{\mathrm{ideal}}$.  Here, $\Sigma I_{\mathrm{foc}}$ is the sum of the intensity values at all the points within the metasurface reflector's focal plane within the beamwidth and $\Sigma I_{\mathrm{ideal}}$ is the sum of all the intensity values at the ideal mirror's focal plane within the beamwidth. This integration approach permits a quantification of actual power instead of intensity at the focus.

\section{\textbf{Conclusion}}
 A broadband metasurface based focusing reflector along with its broadband behavior study has been presented and compared with an ideal mirror. The metasurface based reflector has a broadband power efficiency of 64.1\%, a focal length of 500~mm, and a focal ($1/e^2$) beam width of 14.6~mm, very nearly at the diffraction limit. A pulse width spread of 2.83\% has been observed and is attributed mostly to the strongly dispersive behavior of one unit cell design of the entire reflector's cohort.  While 2.83\% pulse broadening does not initially seem significant, this particular design was selected to minimize dispersion, meaning other design options resulted in worse broadening. Additionally, the GDD analysis reveals that the GDD observed at the focus is dependent both on the dispersive nature of the individual unit cells and superposition of the pulse reflected from all different points on the reflector. It is feasible that design tradeoffs that reduce the focal length, improve efficiency, or increase operating bandwidth could dramatically increase GDD.  In such cases, GDD, if left uncorrected, could prove to be a limiting factor in the achievable data rate of the  metasurface, when it is used in high-bit-rate communications. In the future, improved methods to design metasurface reflectors with managed dispersion can be investigated. Additionally, there is clearly value in additional trade-off studies among the various parameters of focusing metasurface reflectors with an eye toward multi-dimensional optimization.

\section*{Funding}
We gratefully acknowledge support from National Aeronautics and Space Administration (NASA Award Number 80NSSC22K0878)  for this work.
\section*{Acknowledgements}
The authors acknowledge Mr. Karl Strecker and Mr. Russ Messenger for their useful feedback on this manuscript.

\section*{Disclosures}
The authors declare no conflict of interests.

\bibliography{sample}

\begin{thebibliography}{10}

\bibitem{Akyl_6G_Beyond(1)}
Ian~F Akyildiz, Ahan Kak, and Shuai Nie.
\newblock 6g and beyond: The future of wireless communications systems.
\newblock {\em IEEE access}, 8:133995--134030, 2020.

\bibitem{Elme_6G_Wire(2)}
Samar Elmeadawy and Raed~M Shubair.
\newblock 6g wireless communications: future technologies and research
  challenges.
\newblock In {\em 2019 international conference on electrical and computing
  technologies and applications (ICECTA)}, pages 1--5. IEEE, 2019.

\bibitem{Kokko_LOS(3)}
Joonas Kokkoniemi, Janne Lehtomäki, and Markku Juntti.
\newblock A line-of-sight channel model for the 100–450 gigahertz frequency
  band.
\newblock {\em EURASIP Journal on Wireless Communications and Networking},
  2021(1):1--15, 2021.

\bibitem{Naga_Adv(4)}
Tadao Nagatsuma, Guillaume Ducournau, and Cyril~C Renaud.
\newblock Advances in terahertz communications accelerated by photonics.
\newblock {\em Nature Photonics}, 10(6):371--379, 2016.

\bibitem{Xing_Propa(5)}
Yunchou Xing and Theodore~S Rappaport.
\newblock Propagation measurement system and approach at 140 ghz-moving to 6g
  and above 100 ghz.
\newblock In {\em 2018 IEEE global communications Conference (GLOBECOM)}, pages
  1--6. IEEE, 2018.

\bibitem{Coshare_2021}
Yunchou Xing and Theodore~S Rappaport.
\newblock Terahertz wireless communications: Co-sharing for terrestrial and
  satellite systems above 100 ghz.
\newblock {\em IEEE Communications Letters}, 25(10):3156--3160, 2021.

\bibitem{Ai_Multi(6)}
Huifang Ai, Qianlong Kang, Wei Wang, Kai Guo, and Zhongyi Guo.
\newblock Multi-beam steering for 6g communications based on graphene
  metasurfaces.
\newblock {\em Sensors}, 21(14):4784, 2021.

\bibitem{Ultra_MIMO_Jornet}
Ian~F Akyildiz and Josep~Miquel Jornet.
\newblock Realizing ultra-massive mimo (1024× 1024) communication in the
  (0.06–10) terahertz band.
\newblock {\em Nano Communication Networks}, 8:46--54, 2016.

\bibitem{P_Marios}
Marios Poulakis.
\newblock Metamaterials could solve one of 6g’s big problems.
\newblock {\em Proceedings of the IEEE}, 110(9):1151--1158, 2022.

\bibitem{Engheta_Meta_Phy(6)}
Nader Engheta and Richard~W Ziolkowski.
\newblock {\em Metamaterials: physics and engineering explorations}.
\newblock John Wiley \& Sons, 2006.

\bibitem{Schurig_Meta_Cloak(7)}
David Schurig, Jack~J. Mock, B.~J. Justice, Steven~A. Cummer, John~B. Pendry,
  Anthony~F. Starr, and David~R Smith.
\newblock Metamaterial electromagnetic cloak at microwave frequencies.
\newblock {\em Science}, 314(5801):977--980, 2006.

\bibitem{Zhong_Radar(8)}
Shuomin Zhong, Wei Jiang, Peipeng Xu, Taijun Liu, Jifu Huang, and Yungui Ma.
\newblock A radar-infrared bi-stealth structure based on metasurfaces.
\newblock {\em Applied Physics Letters}, 110(6):063502, 2017.

\bibitem{Ling_Negat(9)}
Fang Ling, Zheqiang Zhong, Renshuai Huang, and Bin Zhang.
\newblock A broadband tunable terahertz negative refractive index metamaterial.
\newblock {\em Scientific reports}, 8(1):1--9, 2018.

\bibitem{Shelby_Experi(10)}
Richard~A Shelby, David~R Smith, and Seldon Schultz.
\newblock Experimental verification of a negative index of refraction.
\newblock {\em science}, 292(5514):77--79, 2001.

\bibitem{Wong_Optical(11)}
Zi~Jing Wong, Yuan Wang, Kevin O’Brien, Junsuk Rho, Xiaobo Yin, Shuang Zhang,
  Nicholas Fang, Ta-Jen Yen, and Xiang Zhang.
\newblock Optical and acoustic metamaterials: superlens, negative refractive
  index and invisibility cloak.
\newblock {\em Journal of Optics}, 19(8):084007, 2017.

\bibitem{Chen_Review(11)}
Hou-Tong Chen, Antoinette~J Taylor, and Nanfang Yu.
\newblock A review of metasurfaces: physics and applications.
\newblock {\em Reports on progress in physics}, 79(7):076401, 2016.

\bibitem{Spatio_Temp_Shaltout}
Amr~M Shaltout, Vladimir~M Shalaev, and Mark~L Brongersma.
\newblock Spatiotemporal light control with active metasurfaces.
\newblock {\em Science}, 364(6441):eaat3100, 2019.

\bibitem{Holloway}
Christopher~L Holloway, Edward~F Kuester, Joshua~A Gordon, John O'Hara, Jim
  Booth, and David~R Smith.
\newblock An overview of the theory and applications of metasurfaces: The
  two-dimensional equivalents of metamaterials.
\newblock {\em IEEE antennas and propagation magazine}, 54(2):10--35, 2012.

\bibitem{Zang_Manip(12)}
Xiaofei Zang, Bingshuang Yao, Lin Chen, Jingya Xie, Xuguang Guo, Alexei~V
  Balakin, Alexander~P Shkurinov, and Songlin Zhuang.
\newblock Metasurfaces for manipulating terahertz waves.
\newblock {\em Light: Advanced Manufacturing}, 2(2):148--172, 2021.

\bibitem{Hossein_Repro(14)}
Seyed~Ehsan Hosseininejad, Kasra Rouhi, Mohammad Neshat, Reza Faraji-Dana,
  Albert Cabellos-Aparicio, Sergi Abadal, and Eduard Alarcón.
\newblock Reprogrammable graphene-based metasurface mirror with adaptive focal
  point for thz imaging.
\newblock {\em Scientific reports}, 9(1):1--9, 2019.

\bibitem{Gu_Effi(15)}
Quanlong Yang, Jianqiang Gu, Dongyang Wang, Xueqian Zhang, Zhen Tian, Chunmei
  Ouyang, Ranjan Singh, Jiaguang Han, and Weili Zhang.
\newblock Efficient flat metasurface lens for terahertz imaging.
\newblock {\em Optics express}, 22(21):25931--25939, 2014.

\bibitem{Lee_Dielres(16)}
Wendy~SL Lee, Rajour~T Ako, Mei~Xian Low, Madhu Bhaskaran, Sharath Sriram,
  Christophe Fumeaux, and Withawat Withayachumnankul.
\newblock Dielectric-resonator metasurfaces for broadband terahertz quarter-and
  half-wave mirrors.
\newblock {\em Optics express}, 26(11):14392--14406, 2018.

\bibitem{Aieta_Aberfree(17)}
Francesco Aieta, Patrice Genevet, Mikhail~A Kats, Nanfang Yu, Romain Blanchard,
  Zeno Gaburro, and Federico Capasso.
\newblock Aberration-free ultrathin flat lenses and axicons at telecom
  wavelengths based on plasmonic metasurfaces.
\newblock {\em Nano letters}, 12(9):4932--4936, 2012.

\bibitem{Tang_Ultrawide(18)}
Tong Cai, Shiwei Tang, Bin Zheng, Guangming Wang, Wenye Ji, Chao Qian, Zuojia
  Wang, Erping Li, and Hongsheng Chen.
\newblock Ultrawideband chromatic aberration-free meta-mirrors.
\newblock {\em Advanced Photonics}, 3(1):016001, 2020.

\bibitem{Wang_Broad}
Hang Wang, Fang Ling, Yuan Zhang, Renshuai Huang, Nianchun Sun, and Bin Zhang.
\newblock Broadband and efficient metasurface for beam bending and
  superresolution focusing.
\newblock {\em Superlattices and Microstructures}, 130:512--518, 2019.

\bibitem{Capsso_Achrom_Visi(20)}
Wei~Ting Chen, Alexander~Y Zhu, Vyshakh Sanjeev, Mohammadreza Khorasaninejad,
  Zhujun Shi, Eric Lee, and Federico Capasso.
\newblock A broadband achromatic metalens for focusing and imaging in the
  visible.
\newblock {\em Nature nanotechnology}, 13(3):220--226, 2018.

\bibitem{Witha_Enhanced_(21)}
Xiaolong You, Rajour~T Ako, Wendy~SL Lee, Mei~Xian Low, Madhu Bhaskaran,
  Sharath Sriram, Christophe Fumeaux, and Withawat Withayachumnankul.
\newblock Terahertz reflectarray with enhanced bandwidth.
\newblock {\em Advanced Optical Materials}, 7(20):1900791, 2019.

\bibitem{Qingqing_achro(22)}
Qingqing Cheng, Meilin Ma, Dong Yu, Zhixiong Shen, Jingya Xie, Juncheng Wang,
  Nianxi Xu, Hanming Guo, Wei Hu, and Shuming Wang.
\newblock Broadband achromatic metalens in terahertz regime.
\newblock {\em Science Bulletin}, 64(20):1525--1531, 2019.

\bibitem{Karl_Comp(7)}
Karl Strecker, Sabit Ekin, and John~F O’Hara.
\newblock Compensating atmospheric channel dispersion for terahertz wireless
  communication.
\newblock {\em Scientific reports}, 10(1):1--8, 2020.

\bibitem{Hill_Disp(8)}
Reginald~J Hill.
\newblock Dispersion by atmospheric water vapor at frequencies less than 1 thz.
\newblock {\em IEEE Transactions on Antennas and propagation}, 36(3):423--430,
  1988.

\bibitem{Mandeh_Exp(9)}
Mahboubeh Mandehgar, Yihong Yang, and D~Grischkowsky.
\newblock Experimental confirmation and physical understanding of ultra-high
  bit rate impulse radio in the thz digital communication channels of the
  atmosphere.
\newblock {\em Journal of Optics}, 16(9):094004, 2014.

\bibitem{Karl_Funda}
Karl Strecker, Sabit Ekin, and John~F O’Hara.
\newblock Fundamental performance limits on terahertz wireless links imposed by
  group velocity dispersion.
\newblock {\em IEEE Transactions on Terahertz Science and Technology},
  12(1):87--97, 2021.

\bibitem{Heath_over(10)}
Robert~W Heath, Nuria Gonzalez-Prelcic, Sundeep Rangan, Wonil Roh, and Akbar~M
  Sayeed.
\newblock An overview of signal processing techniques for millimeter wave mimo
  systems.
\newblock {\em IEEE journal of selected topics in signal processing},
  10(3):436--453, 2016.

\bibitem{Sari_over(11)}
Hadi Sarieddeen, Mohamed-Slim Alouini, and Tareq~Y Al-Naffouri.
\newblock An overview of signal processing techniques for terahertz
  communications.
\newblock {\em Proceedings of the IEEE}, 2021.

\bibitem{Farrell_Perf(12)}
Somayeh Mohammady, Ronan Farrell, David Malone, and John Dooley.
\newblock Performance investigation of peak shrinking and interpolating the
  papr reduction technique for lte-advance and 5g signals.
\newblock {\em Information}, 11(1):20, 2019.

\bibitem{Tae_N20}
Gyeong-Ryul Kim, Hwa-Bin Lee, and Tae-In Jeon.
\newblock Terahertz time-domain spectroscopy of low-concentration n 2 o using
  long-range multipass gas cell.
\newblock {\em IEEE Transactions on Terahertz Science and Technology},
  10(5):524--530, 2020.

\bibitem{Tae_water}
Jae-Gwang Kwon, Mun-Won Park, and Tae-In Jeon.
\newblock Determination of the water vapor continuum absorption by thz pulse
  transmission using long-range multipass cell.
\newblock {\em Journal of Quantitative Spectroscopy and Radiative Transfer},
  272:107811, 2021.

\bibitem{Azad_Meta_MW}
Sinhara~R Silva, Abdur Rahman, Wilton de~Melo Kort-Kamp, Jeremiah~J Rushton,
  John Singleton, Antoinette~J Taylor, Diego~AR Dalvit, Hou-Tong Chen, and
  Abul~K Azad.
\newblock Metasurface-based ultra-lightweight high-gain off-axis flat parabolic
  reflectarray for microwave beam collimation/focusing.
\newblock {\em Scientific reports}, 9(1):1--7, 2019.

\bibitem{Goodman}
Joseph~W Goodman.
\newblock {\em Introduction to Fourier optics}.
\newblock Roberts and Company publishers, 2005.

\bibitem{Oppenheim}
Alan~V. Oppenheim, Alan~S. Willsky, and S.~Hamid Nawab.
\newblock Time and frequency characterization of signals and systems.
\newblock In {\em Signals \& Systems}, chapter~6, page 432. Prentice {H}all,
  Upper Saddle River, New Jersey 07458, USA, 2nd edition, 1996.

\bibitem{Russ}
Russ Messenger, Karl Strecker, Sabit Ekin, and John~F O'Hara.
\newblock Dispersion from diffuse reflectors and its effect on terahertz
  wireless communication performance.
\newblock {\em IEEE Transactions on Terahertz Science and Technology},
  11(6):695--703, 2021.

\end{thebibliography}
\bibliographystyle{unsrt}  

\end{document}